\documentclass[iop,apj,twocolumn]{emulateapj}

\usepackage{natbib}
\usepackage{amsfonts}
\usepackage{amsmath}
\usepackage{amssymb}
\usepackage{bm}
\usepackage{dcolumn}
\usepackage{graphicx}
\usepackage{graphics}
\usepackage[latin1]{inputenc}
\usepackage{latexsym}
\usepackage{rotating}
\usepackage[colorlinks=false]{hyperref}
\usepackage[all]{hypcap}
\usepackage{xspace} 
\usepackage[usenames]{color}
\usepackage{mathrsfs}

\usepackage{ulem}
\definecolor {darkgreen}{rgb}{0.2,.9,0.2}

\newcommand\be{\begin{equation}}
\newcommand\ba{\begin{eqnarray}}
\newcommand\ee{\end{equation}}
\newcommand\ea{\end{eqnarray}}
\newcommand\bw{\begin{widetext}}
\newcommand\ew{\end{widetext}}

\newcommand{\nn}{\nonumber}

\newcommand{\BH}{{\mbox{\tiny BH}}}

\submitted{Received 2014 February 17; accepted 2014 April 19; published 2014 May 16}
\journalinfo{The Astrophysical Journal, 788:15 (5pp), 2014 June 10
  \hfill doi:\href{http://dx.doi.org/10.1088/0004-637X/788/1/15}{10.1088/0004-637X/788/1/15}}

\begin{document}
\title{Three-Hair Relations for Rotating Stars: Nonrelativistic Limit}

\author{Leo C. Stein\altaffilmark{1,3},
Kent Yagi\altaffilmark{2},
and Nicol\'as Yunes\altaffilmark{2}}
\affiliation{${}^1$ Center for Radiophysics and Space Research, Cornell
  University, Ithaca, NY 14853 USA;
\href{mailto:leostein@astro.cornell.edu}{leostein@astro.cornell.edu}\\
${}^2$ Department of Physics, Montana State University, Bozeman, MT 59717, USA}
\altaffiltext{3}{Einstein fellow.}

\begin{abstract}
The gravitational field outside of astrophysical black holes is
completely described by their mass and spin frequency, as expressed by
the no-hair theorems. These theorems assume vacuum spacetimes, and thus
they apply only to black holes and not to stars. Despite this, we analytically
find that the gravitational potential of arbitrarily rapid rigidly rotating stars
can still be described completely by only their mass, spin angular momentum, and
quadrupole moment. Although these results are obtained in the nonrelativistic limit
(to leading order in a weak-field expansion of general relativity, GR), they are also consistent with
fully relativistic numerical calculations of rotating neutron stars. This description of 
the gravitational potential outside the source 
in terms of just three quantities is approximately universal
(independent of equation of state). Such universality may be used to break degeneracies
in pulsar and future gravitational wave observations to extract more physics
and test GR in the strong-field regime.
\end{abstract}
\maketitle

\section{Introduction}
Neutron stars (NSs) are excellent laboratories
to study extreme, relativistic astrophysics at
supra-nuclear densities~\citep{lattimer-prakash-review}. NS
observations can also help us carry out more stringent tests of general relativity (GR) in
the strong gravity regime~\citep{lrr-2006-3}. These observations are currently limited to binary
pulsar measurements, but soon gravitational wave (GW) observations of NSs
may allow astrophysical and fundamental physics studies~\citep{lrr-2013-9}.

To fully exploit NS observations, one needs to understand how
the NS's interior imprints onto observables. The nuclear
equation of state (EoS; the relation between the fluid's state
variables, e.g.,~pressure and energy density) determines the radial
density profile and the compressibility of the star. Thus, the EoS
connects microscopic physics to macroscopic observables, like the mass
and radius.


The EoS also controls the NS's shape, which is aspherical due to
rotation. The stellar shape affects the gravitational field outside the source, which
controls various NS observables. The exterior gravitational field can
be described through a multipolar decomposition~\citep{Backdahl:2005uz,Backdahl:2006ed}, 
just as when describing the exterior electromagnetic field of a charged object with
multipole moments.
 
The NS's multipole moments enter directly into astrophysical
observables. Specifically, the $\ell=0$ multipole moment (monopole)
corresponds to the mass. The $\ell=1$ moment (mass-current dipole)
corresponds to the star's spin angular momentum, the product
of the star's moment of inertia $I$ about its spin axis and its spin angular 
frequency $\Omega$. The $\ell = 2$ moment (mass quadrupole $Q$) 
can be mapped to the quadrupolar deformation of the star~\citep{hartle1967,baubock}. 
NS observations, such as X-ray atomic line profiles from NS surfaces~\citep{2002Natur.420...51C}, depend on
the star's EoS-dependent multipole moments~\citep{Baubock:2012bj}.


Na\"ively, one may be discouraged by needing to know an infinite
number of EoS-dependent multipole moments to describe the 
gravitational field outside NSs.  Recently, however, approximately EoS-independent
relations were discovered between certain quantities related to the
star's multipole moments, i.e.,~between the moment of inertia ($I$), 
the tidal Love number, and the quadrupole moment ($Q$) 
of slowly rotating NSs~\citep{I-Love-Q-Science,I-Love-Q-PRD}.  Both $I$
and $Q$ depend strongly on the EoS when written as functions of NS
mass or compactness, but $I$ as a function of $Q$ is approximately
EoS-independent.

Such approximately EoS-independent relations are of widespread interest to a variety of communities because they describe the gravitational field outside a NS without knowledge of the star's EoS~\citep{I-Love-Q-Science,I-Love-Q-PRD}.  The I-Love-Q relations can
be used in X-ray observations of millisecond pulsars~\citep{Psaltis:2013fha}, as well as in 
GW observations of NS mergers~\citep{I-Love-Q-Science,I-Love-Q-PRD}. 
In the latter, these relations may break degeneracies between the star's spin angular momentum 
and its quadrupolar deformation, allowing for better
measurements of both, and for EoS independent tests of
GR~\citep{kramer-double-pulsar,I-Love-Q-Science,I-Love-Q-PRD}.

Most of these applications require observations of old NSs, which are expected to 
have moderate to weak magnetic fields and slow rotation. Nonetheless, the universal
$I$-Love-$Q$ relations have recently been extended to large tidal
deformations~\citep{maselli}, moderate magnetic
fields~\citep{I-Love-Q-B}, and rapidly rotating
stars~\citep{doneva-rapid,Pappas:2013naa,Chakrabarti:2013tca}. A minor
controversy recently arose regarding rotation. We here
analytically resolve this controversy to show that the $I$--$Q$
relations remain almost EoS-independent for rapidly rotating stars,
supporting the numerical work of~\citet{Pappas:2013naa} and \citet{Chakrabarti:2013tca}. 
We henceforth focus on weakly magnetized and cold stars, but with
arbitrarily rapid,
rigid rotation.

The existence of approximately EoS-independent $I$--$Q$ relations
suggests that more general relations may exist for other
multipole moments. Such an idea is reminiscent of the black hole (BH)
no-hair theorems, which state that their
gravitational field outside the horizon is completely described by their mass
(the $\ell=0$ mass moment) and spin (the $\ell=1$ mass-current moment;~\citet{israel,hawking-uniqueness}).
However, the no-hair theorems assume pure vacuum spacetimes,
and thus do not apply to NSs.
Despite this, we show that the multipole moments that describe the gravitational
field outside an arbitrarily rapid, rigidly rotating NS can be completely described in an
approximately EoS-independent way by three moments: the mass monopole (mass), the
mass--current dipole (angular momentum), and the mass quadrupole.
We work in the nonrelativistic limit of gravity, i.e.,~to leading
``Newtonian'' order in a weak-field/slow-motion expansion~\citep{Blanchet:2002av}.

\section{Multipole Moments}
\label{sec:multipole-moments}
In GR, a stationary, asymptotically flat, and axisymmetric spacetime outside a source is completely described by its mass and mass--current multipole moments. To leading ``Newtonian'' order
in a weak-field/slow-motion expansion, the mass and mass--current moments of a rigidly rotating star are given by~\citep{Ryan:1996nk}
\begin{align}
\label{eq:M-ell-integ}
M_\ell &= 2 \pi \int^\pi_0 \int^{R_*(\theta)}_0 \!\!\! \rho(r,\theta) \;  P_\ell (\cos \theta) \; \sin\theta  d\theta \; r^{\ell +2} dr\,, 
\\
\label{eq:S-ell-integ}
S_\ell &= \frac{4 \pi \Omega}{\ell+1}  \int^\pi_0 \int^{R_*(\theta)}_0 \!\!\! \rho(r,\theta) \frac{d P_\ell (\cos \theta)}{d\cos\theta}  \;  \sin^3\theta  d\theta \; r^{\ell +3} dr\,,
\end{align}
where $R_*(\theta)$ is the stellar surface profile in spherical coordinates, 
$\rho(r,\theta)$ is the mass density, $\Omega$ is the stellar spin angular velocity, and $P_\ell(\cos{\theta})$ are Legendre polynomials 
(we work in geometric units, $G = 1 = c$). In Newtonian gravity, 
the mass moments are sufficient to fully describe the gravitational potential 
of axially symmetric stars, since the mass--current moments are higher-order in $v/c = {\cal{O}}(R_{*} \Omega/c) \ll 1$
(momentarily restoring the factors of $c$). Since astrophysical stars are reflection-symmetric about the equator,
$M_{2\ell+1}=0=S_{2\ell}$. These moments are algebraically related to, but distinct
from, the Geroch--Hansen (GH) moments~\citep{Geroch:1970cc,Geroch:1970cd,hansen:46,Gursel} and the
Thorne moments~\citep{thorne-MM} in the nonrelativistic limit~\citep{pappas-apostolatos}. 
In this limit, the $\ell$th GH and Thorne moments are
equivalent~\citep{Gursel} to our moments
(Equations~\eqref{eq:M-ell-integ} and \eqref{eq:S-ell-integ})
\citep{Gurlebeck:2012mb}.

Three multipole moments will here play a special role: $M_{0}$, $S_{1}$, and $M_{2}$. 
The mass monopole is simply the total mass, i.e.,~$M_{0} = M$. 
The mass--current dipole $S_{1} = I \Omega$
is the magnitude of the star's spin angular momentum. 
The mass quadrupole $M_{2}$
is related to the quadrupole tensor, the former being proportional to
the contraction of the latter with two copies of the spin axis unit vector.

Without further simplifications, the above integrals are impossible to
compute analytically. Numerical calculations are possible, but these
lack the insight that analytical results yield. To make
analytic progress, we adopt the elliptical isodensity
  approximation of~\citet{Lai:1993ve}: (1) we
treat surfaces of constant density as self-similar ellipsoids of given
ellipticity, and (2) the density profile in terms of the isodensity
radius is identical to that of a spherically symmetric star of the
same volume as the rotating star.
In a realistic star, the isodensity contours are more spherical
toward the center and more oblate toward the surface. Despite this, 
this approximation has been shown to
be extremely accurate, with errors on the lowest multipole moments
of at most $3\%$ relative to full
numerical calculations at rotation frequencies that saturate the mass--shedding limit~\citep{Lai:1993ve}.
This difference is unimportant for the calculation of multipole moments, because
the integrands at larger radii contribute more than in the inner core~\citep{I-Love-Q-PRD}.

Let us introduce a suitable coordinate system adapted to the
isodensity approximation $x^{i} = \tilde{r}\Theta(\cos{\theta})
n^{i}$, where $n^{i} = (\sin{\theta} \cos{\phi},\sin{\theta}
\sin{\phi}, \cos{\theta})$ is the unit direction vector, with
\begin{equation}
\Theta(\cos\theta) \equiv \sqrt{\frac{1-e^2}{1-e^2 (1-\cos^2\theta)}}\,,
\end{equation}
where $e = \sqrt{1- a_{3}^2/a_{1}^2}$ is the star's eccentricity and
$\tilde{r}={\rm{const.}}$ are isodensity surfaces.  The stellar surface, 
at $\tilde{r} = a_{1}$, is an oblate ellipsoid with semi-major and
semi-minor axes $a_{1}$ and $a_{3}$, and geometric mean radius 
$R = (a_{1}^{2} a_{3})^{1/3} = a_{1} (1 - e^{2})^{1/6}$.

Using this coordinate system, the angular and radial integrals in
Equations~\eqref{eq:M-ell-integ} and \eqref{eq:S-ell-integ} can be
separated into
\begin{align}
\label{eq-for-M}
M_{\ell}  ={}& 2\pi \; I_{\ell,3} \; R_{\ell} \,,\\
\label{eq-for-S}
S_\ell ={}& \frac{4 \pi \ell}{2 \ell +1} \Omega
\left(I_{\ell-1,5}-I_{\ell+1,3}\right) R_{\ell+1}\,,
\end{align}
where we have used Legendre polynomial identities and defined
\begin{align}
R_{\ell}  \equiv{}& \int_{0}^{a_{1}} \!\!\! \rho(\tilde{r}) \tilde{r}^{\ell+2} d\tilde{r}\,,
&
I_{\ell,k}  \equiv{}& \int_{-1}^{+1} \!\!\! \Theta(\mu)^{\ell+k} P_{\ell}(\mu) d\mu\,,
\end{align}
with $\mu=\cos\theta$. 

The integrals $I_{\ell,3}$ and $I_{\ell,5}$ can be done in closed form.
Using Equation~(7.226.1) from~\citet{2007tisp.book.....G} and deriving a
related identity,
we find
\begin{align}
\label{eq:I-for-M}
I_{\ell,3} ={}&  (-)^{\frac{\ell}{2}}\frac{2 }{\ell+1}\sqrt{1-e^2} e^\ell\,, \\
\label{eq:I-for-S}
I_{\ell-1,5}-I_{\ell+1,3}={}& (-)^{\frac{\ell-1}{2}}\frac{2 (2 \ell+1) }{\ell (\ell+2)}\sqrt{1-e^2}e^{\ell-1}\,.
\end{align}
Equations~\eqref{eq:I-for-M} and~\eqref{eq:I-for-S} are only 
evaluated for even and odd $\ell$, respectively, so they are both real.

All of the EoS dependence is in $\Omega(e)$ 
and $R_{\ell}$, with the latter containing the radial density profile.
Realistic EoSs may be parameterized by a piecewise collection
of polytropes, each of the form $P=K\rho^{1+1/n}$~\citep{Read:2008iy,lattimer-prakash-review}.
For simplicity, we here consider single polytropes, but our
results are extendable to piecewise polytropes. Let us
transform to dimensionless variables
following the Lane--Emden approach~\citep{2004sipp.book.....H}.
Take $\rho = \rho_{c} \left[\vartheta(\xi)\right]^{n}$\,,
where $\rho_{c} = M/(4 \pi R^{3}) \xi_{1}/|\vartheta'(\xi_{1})|$
is the central density, with $M$ the stellar mass, $\vartheta(\xi)$
a dimensionless function related to density, and
$\xi = (\xi_{1}/a_{1}) \tilde{r}$ a dimensionless radius,
such that $\xi = \xi_{1}$ is the stellar surface.
The radial integral then becomes 
$R_{\ell} = \rho_{c} \left({a_{1}}/{\xi_{1}}\right)^{\ell + 3} {\cal{R}}_{n,\ell}$,
where
\be
{\cal{R}}_{n,\ell} =  \int_{0}^{\xi_{1}}
 \left[\vartheta_{\rm sph}(\xi)\right]^{n} \xi^{\ell+2} d \xi\,,
\ee
and we used the elliptical isodensity approximation to replace $\vartheta$ by its
spherically symmetric counterpart $\vartheta_{\rm sph}$, a Lane--Emden function.

Putting it all together, we find
\begin{align}
\label{eq:M2l2-CM}
M_{2 \ell + 2} ={}&
\frac{\left(-\right)^{\ell+1}}{2 \ell + 3}
\frac{e^{2\ell+2}}{(1 - e^{2})^{\frac{\ell+1}{3}}}
\frac{{\cal{R}}_{n,2+2 \ell}}{\xi_{1}^{2 \ell + 4} |\vartheta'(\xi_{1})|}
\frac{M^{2 \ell+3}}{C^{2\ell+2}}\,,
\\
\label{eq:S2l1-CM}
S_{2 \ell + 1} ={}&
\frac{(-)^{\ell}}{2 \ell + 3}
\frac{2 \Omega \; e^{2 \ell}}{(1 - e^{2})^{\frac{\ell+1}{3}}}
\frac{{\cal{R}}_{n,2+2\ell}}{\xi_{1}^{2 \ell + 4} |\vartheta'(\xi_{1})|}
\frac{M^{2 \ell+3}}{C^{2\ell+2}}\,.
\end{align}
These expressions are valid for all rotation periods. The relation between eccentricity
and angular frequency is given in the elliptical isodensity approximation by~\citep{Lai:1993ve}
\begin{align}
\label{eq-Omega-of-e}
\Omega(e) ={}&
\frac{3}{2} \xi_{1}^{2}
\left[\frac{C^{3}  |\vartheta'(\xi_{1})|}{(5-n) M^{2} {\cal{R}}_{n,2}}\right]^{1/2} f(e)\,,
\end{align}
with 
\begin{align}
f(e) ={}& \left[ - 6 e^{-2} \left(1 - e^{2}\right)
\right.
\nn \\
& \left.
{}+2 e^{-3} \left(1 - e^{2}\right)^{1/2} \left(3 - 2 e^{2}\right) \arcsin{(e)} \right]^{1/2}\,,
\end{align}
where $C = M/R$ is the stellar compactness.
In the $n=0$ and $n=1$ cases,  one can calculate these moments exactly and purely analytically. 

\section{Universality and Breakdown}
Let us first work with the dimensionless moments
\renewcommand{\bar}[1]{\overline{#1}}
\be
\label{eq:def-dimensionless}
\bar{M}_{\ell} = (-)^{\frac{\ell}{2}} \frac{M_{\ell}}{M^{\ell + 1}
  \chi^{\ell}}\,, \quad \bar{S}_{\ell} = (-)^{\frac{\ell-1}{2}}
\frac{S_{\ell}}{M^{\ell + 1} \chi^{\ell}}\,, \ee where $\chi \equiv
S_{1}/M^{2}$. 
With this normalization, $\bar{M}_0 = 1$ and $\bar{S}_1 = 1$ always, and BHs
have $\bar{M}_{2\ell}^\BH = 1 = \bar{S}_{2\ell+1}^\BH$.

\begin{figure}[t]
\centering
\includegraphics[width=\columnwidth,clip=true]{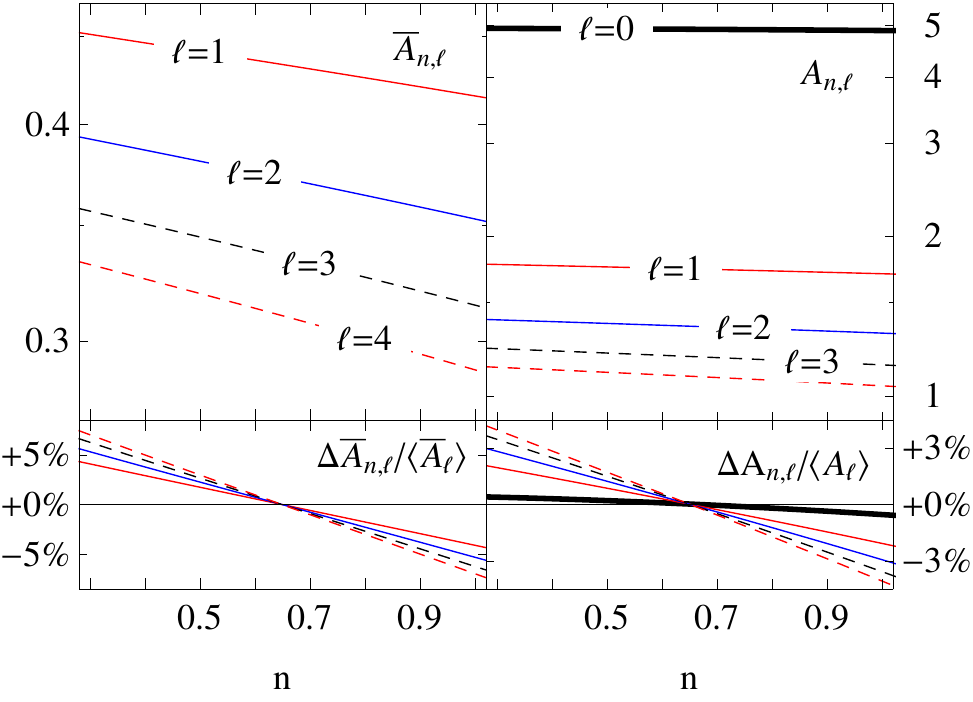}
\caption{
Universality of coefficients $\bar{A}_{n,\ell}$
(left) and $A_{n,\ell}$ (right) with polytropic index $n$.
Top panels give the coefficients themselves, while the bottom panels show the
relative fractional difference between the coefficients and the
averaged value over $n\in [0.3,1]$.
The fractional variation in $A_{n,0}$ (the $M_{2}$--$S_{1}$ relation)
is less than 0.5\% over this range. Even up to $\ell=4$ (which controls the
$M_{10}$--$S_{9}$ relation), there is less than 5\% variation.
\label{fig:A-Abar-combined}
}
\end{figure}

Three-hair NS relations can be obtained as follows.
First, from Equations~\eqref{eq-for-M} and~\eqref{eq-for-S}, we find 
\be
\label{eq:1st-no-hair}
\bar{M}_{2 \ell + 2} = \bar{M}_{2} \; \bar{S}_{2 \ell + 1}\,.
\ee
Note that this relation holds independently of the rotation period and the EoS.
Second, we eliminate $C$ from Equation~\eqref{eq:M2l2-CM} by using Equation~\eqref{eq:S2l1-CM}, where $\Omega$ is eliminated by using $S_1$. This gives
\be
\label{eq:2nd-no-hair}
\bar{M}_{2 \ell + 2} = \bar{A}_{n,\ell} \; (\bar{S}_{2 \ell+1})^{1+1/\ell}\qquad (\ell>0)\,,
\ee
which again holds independently of the rotation period, but depends on the EoS through the coefficients
\begin{equation}
\label{eq:bar-A-def}
\bar{A}_{n,\ell} =
\frac{  (2 \ell +3)^{1/\ell}}{3^{(1+1/\ell)}}
\frac{{\cal{R}}_{n,2}^{1+1/\ell}{\cal{R}}_{n,2+2 \ell}^{-1/\ell}}{{|\vartheta'(\xi_{1})| \; \xi_{1}^{2}}}\,.
\end{equation}
These coefficients have a small variability with $n$, as
seen in the left panels of Figure~\ref{fig:A-Abar-combined}.

\begin{figure}[tb]
\centering
\includegraphics[width=\columnwidth,clip=true]{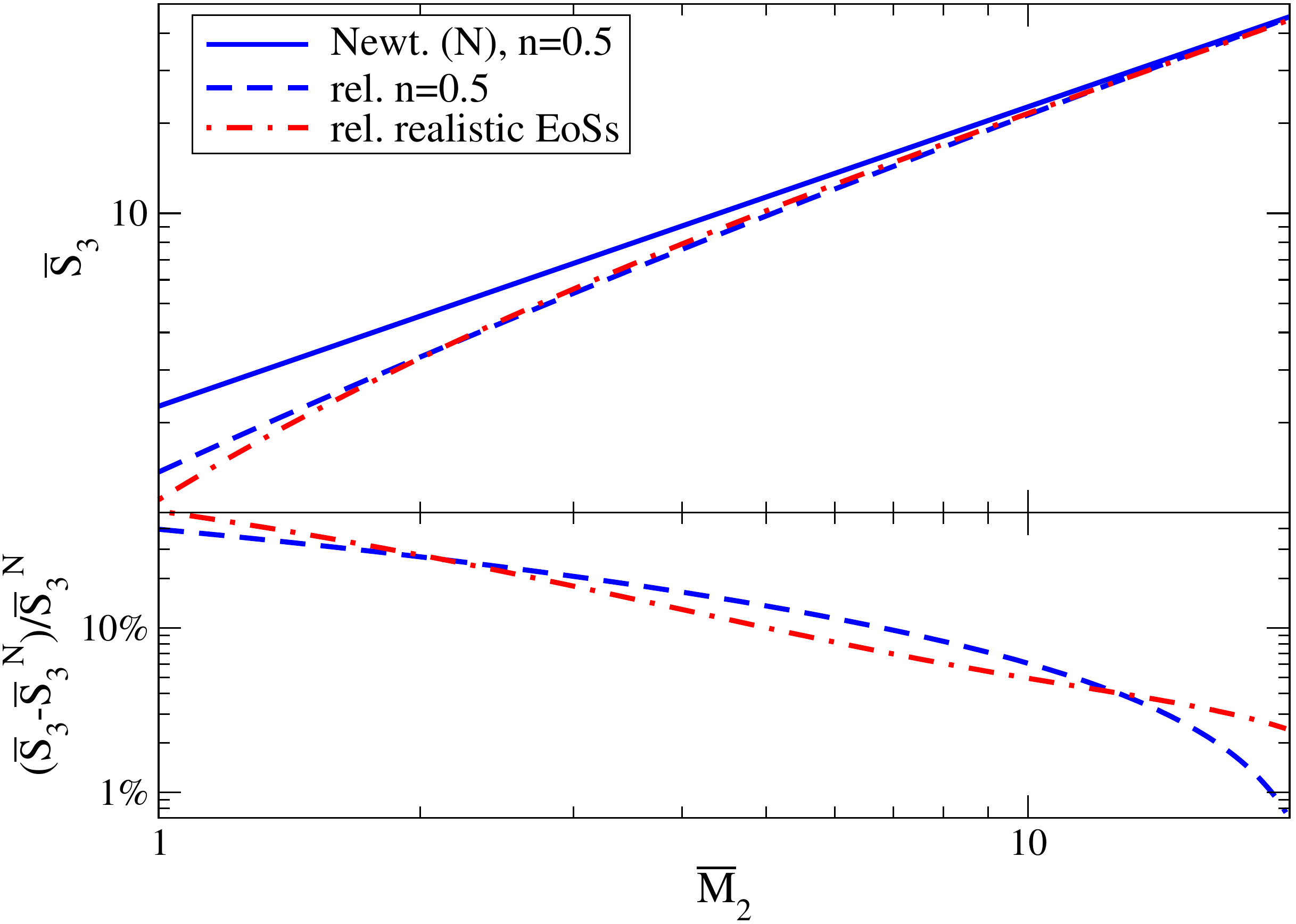}
\caption{%
  Comparison between relativistic and weak-field
  three-hair relations. 
 Top: weak-field (solid blue, this work), and relativistic $\bar{S}_3$
  vs.~$\bar{M}_2$ relation for an $n=0.5$ polytrope (blue dashed) 
 and the fit among realistic EoSs (red dot-dashed) from~\citet{Pappas:2013naa}.
 The parameter along all curves is stellar
  compactness. Bottom: fractional difference between the relativistic
  and our weak-field results.
  \label{fig:comparison-with-pappas}}
\end{figure}

We can now use Equations~\eqref{eq:1st-no-hair} and~\eqref{eq:2nd-no-hair} to find three-hair relations for all multipoles in terms of the first three:
\be
\label{eq:3-hair}
\bar{M}_{2\ell+2} + i \bar{S}_{2\ell+1} = \bar{B}_{n,\ell}  \bar{M}_2^{\ell} (\bar{M}_2 + i \bar{S}_1)\,,
\ee
where $\bar{B}_{n, \ell} = (\bar{A}_{n,\ell})^{-\ell}$. Note that
these relations are independent of the rotation period, which 
analytically supports the fully numerical
results of~\citet{Pappas:2013naa} in GR for the $\ell=1$ case
(see also Figure~\ref{fig:comparison-with-pappas} and its accompanying discussion).

Converting back to dimensional
moments through Equation~\eqref{eq:def-dimensionless}, we find
\be
\label{eq:3-hair-dimensional}
M_{\ell} + i \frac{q}{a} S_{\ell} = \bar{B}_{n,\lfloor \frac{\ell-1}{2} \rfloor}  M (i q)^\ell\,,
\ee
where $a \equiv S_1/M$, $iq \equiv \sqrt{M_2/M}$, and $\lfloor x
\rfloor$ denotes the largest integer not exceeding $x$. Note
that although Equation~\eqref{eq:bar-A-def} is only valid for $\ell>0$,
the relation in Equation~\eqref{eq:3-hair-dimensional} is also valid when $\ell=0$ and $\ell=1$
with $\bar{B}_{n,-1}=1=\bar{B}_{n,0}$, where 
the value of $\bar{B}_{n,-1}$ is obtained through the Lane-Emden equation.
Equation~\eqref{eq:3-hair-dimensional} resembles the
BH no-hair relation, $M_{\ell}^{\BH} + i
S_{\ell}^{\BH} = M (i a)^\ell$~\citep{hansen:46}.

Figure~\ref{fig:comparison-with-pappas} compares our nonrelativistic/weak-field 
three-hair relations (Equation~\eqref{eq:3-hair} with $\ell=1$) to the numerical results
in full GR of~\citet{Pappas:2013naa} for an $n=0.5$ polytropic EoS, as well as to
their analytic fit to data for 10 realistic EoSs, with a fitting error of $\mathcal{O}(1\%)$. 
The results of~\citet{Pappas:2013naa} are obtained by numerically solving the GR 
equations of structure for NSs. Observe that their relativistic results match our weak-field 
relation as $C$ decreases ($\bar{M}_2$ increases), approaching the weak-field regime.
The agreement in the low-compactness regime is better than 3\%,
which is comparable to the accuracy of the elliptical isodensity approximation
to the lowest multipole moment~\citep{Lai:1993ve}. 
Even for very compact stars, 
as $\bar{M}_2 \rightarrow 1$, our results differ from the fully relativistic one 
by roughly 40\% at most.  Observe also that our \emph{single-polytropic} results are 
consistent with the full GR results that use realistic EoSs.

One may believe that the elliptical isodensity 
approximation reduces the number of degrees of freedom of the stellar models, 
and this is why the relations between multipoles are approximately EoS-independent.
However, \citet{Pappas:2013naa} found such universality without imposing
this approximation, and their results are consistent with ours.
The $\mathcal{O}(1\%)$ EoS-variability found in~\citet{Pappas:2013naa}
is consistent with the variation in $\bar{A}_{n,1}$ in Figure~\ref{fig:A-Abar-combined}.

Let us now take the slow-rotation limit, keeping terms to leading-order in $M\Omega \ll 1$ and $e \ll1$.
Expanding Equations~\eqref{eq:M2l2-CM}--\eqref{eq-Omega-of-e} and eliminating $C$ in favor
of $S_{2\ell+1}$, we find
\be
\label{eq:2nd-no-hair-new}
|M_{2 \ell + 2}| =
A_{n,\ell} \left[ \frac{|S_{2\ell+1}|^{5(\ell+1)}}{M^{\ell+4} (M \Omega)^{\ell+1}} \right]^{\frac{1}{5\ell+2}}
\left[1 + {\cal{O}}(M \Omega)^{2}\right]\,,
\ee
with the dimensionless coefficients
\be
\label{eq:A-def}
A_{n,\ell} =
\left\{ \left[\frac{25\left(5 - n\right)^2}{1152} \right]^{\ell + 1}
\frac{(2 \ell + 3)^{3}{\cal{R}}_{n,2}^{2 \ell + 2}}{{\cal{R}}_{n,2\ell+2}^{3} \xi_{1}^{2 \ell -4}|\vartheta'(\xi_{1})|^{2 \ell -1}}
\right\}^{\frac{1}{5 \ell + 2}} \!\!\!\!\!\!\!\!\,.
\ee
These coefficients have even smaller variation with $n$
than $\bar{A}_{n,\ell}$ (Figure~\ref{fig:A-Abar-combined}). 
In the slow-rotation limit, $R$ and therefore $C$ are constants
independent of $\Omega$, since $R = a_{1} + {\cal{O}}(M \Omega)^{2}$.
For an $n=0$ polytrope, Equation~\eqref{eq:2nd-no-hair-new} reproduces exactly 
the leading-order, weak-field expansion of the $I$--$Q$ relation 
of~\citet{I-Love-Q-PRD}, obtained without imposing the 
elliptical isodensity approximation.
For an $n=1$ polytrope, the coefficient of the $I$--$Q$ relation differs by 3\%, which
is consistent with the validity of the approximation.

Let us now return to the $I$--$Q$ relation for arbitrary spin in order
to address the minor controversy between the results
of~\citet{doneva-rapid}, \citet{Pappas:2013naa} and
\citet{Chakrabarti:2013tca}.
Equations~\eqref{eq:M2l2-CM}--\eqref{eq-Omega-of-e} with $\ell=0$
imply
\be
\bar{M}_2 = \frac{\bar{I}}{2} \frac{e^2}{\chi^2}\,, \quad 
\chi = \frac{\sqrt{15}}{4} \frac{\bar{I}^{1/4}}{A_{n,0}^{1/2}} \frac{f(e)}{(1-e^2)^{1/4}}\,,
\ee
where recall that $\chi = S_{1}/M^{2}$ and $\bar{I} = \chi/M \Omega$.
Although $\chi$ diverges as $e \to 1$,
this is an unphysical limit (ellipsoids become degenerate in that limit), and the 
currently observed pulsars~\citep{2006Sci...311.1901H} all have $e \lesssim 0.7$. To find a relation
for $\bar{M}_{2}$ in terms of $\bar{I}$ and $\chi$, one needs to invert the
expression for $\chi$ to obtain $e (\chi, \bar{I}, n)$. For any given $\chi$, however,  
the $\bar{M}_{2}$--$\bar{I}$ relation will depend on the EoS only through $A_{n,0}$, 
regardless of the magnitude of $\chi$. Since $A_{n,0}$ is approximately EoS-independent, 
the $\bar{M}_{2}$--$\bar{I}$--$\chi$ relation is as well, confirming the
results of~\citet{Pappas:2013naa} and \citet{Chakrabarti:2013tca} in
full GR. Moreover, one
finds that the $I$--$Q$ relation is approximately EoS-independent both for fixed $\chi$ or fixed 
$M\Omega$, confirming the results of~\citet{Chakrabarti:2013tca}, and
resolving this controversy.

\begin{figure}[tb]
  \centering
  \includegraphics[width=\columnwidth]{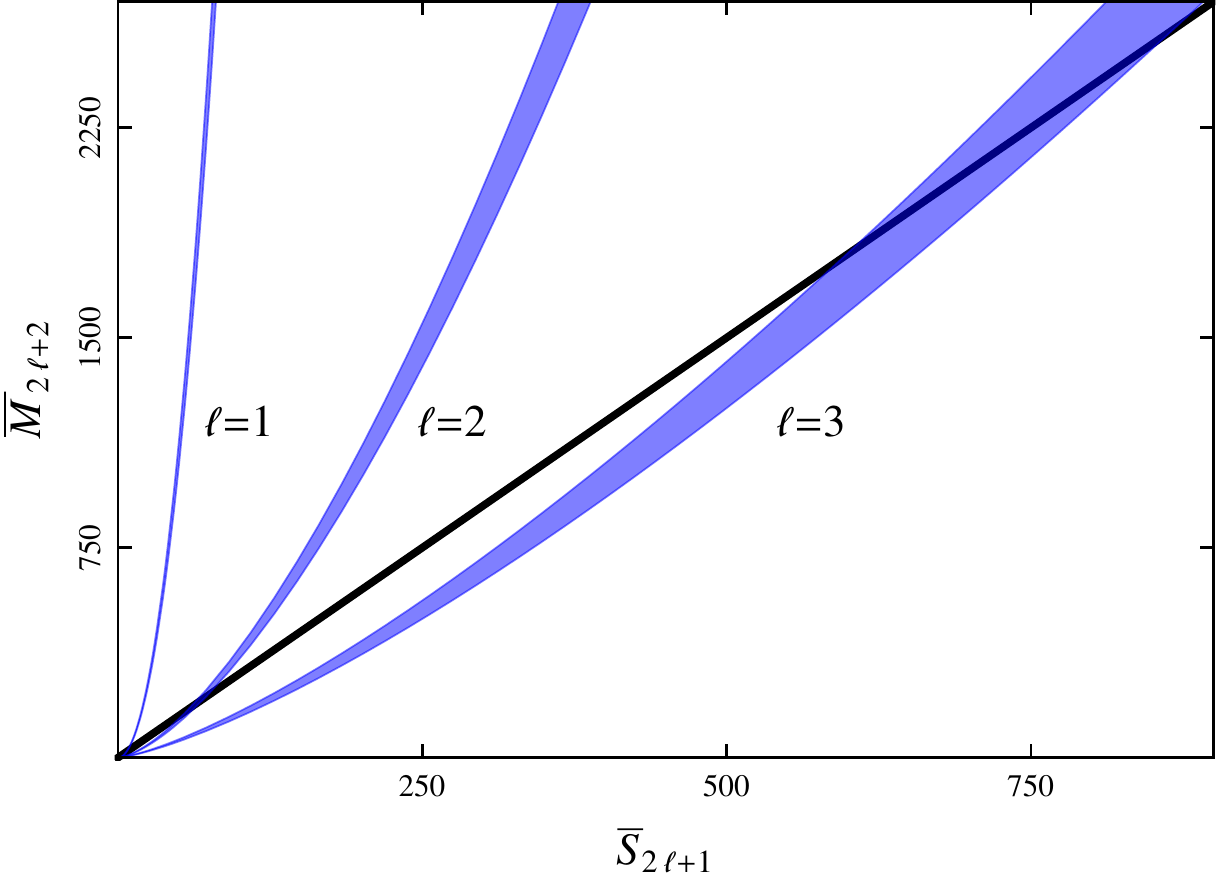}
  \caption{
    Schematic breakdown of universality at high $\ell$.
    A solution for $\bar{M}_{2\ell+2},\bar{S}_{2\ell+1}$ comes from
    the intersection of a straight line (Equation~\eqref{eq:1st-no-hair};
    here plotted with $\bar{M}_{2}=3$)
    with a power law [Equation~\eqref{eq:2nd-no-hair}]. The width of the
    power law comes from the EoS dependence of the prefactor
    $\bar{A}_{n,\ell}$. At high $\ell$, the power law approaches
    linearity and the universality degenerates, as the intersection
    region grows.
  \label{fig:growing-degeneracy}
}
\end{figure}

Finally, although the relations in Equations~\eqref{eq:1st-no-hair}, \eqref{eq:2nd-no-hair}
and~\eqref{eq:2nd-no-hair-new} hold for arbitrary $\ell$, the
decoupling used to obtain Equation~\eqref{eq:3-hair} breaks for large
$\ell$. This is because for large $\ell$, Equation~\eqref{eq:2nd-no-hair} approaches
linearity in $S_{2 \ell + 1}$, and thus the solution becomes degenerate, as shown
in Figure~\ref{fig:growing-degeneracy}.
Observe how the power-law approaches linearity as $\ell$ increases, and the
intersection region with the straight line grows.

\section{Conclusions}
We found approximately EoS-independent relations
for all weak-field multipole moments of arbitrarily rapidly, rigidly rotating stars
in terms of the first three: the stellar mass (monopole),
the angular momentum (mass--current dipole) and the quadrupole moment. This universality
is found for polytropic EoSs with $n \in (0.5$--$1.0)$, a range of
single-polytrope proxy models which are most like
NSs~\citep{lattimer-prakash2001,flanagan-hinderer-love}. This
approximate universality is valid for arbitrary rotation, resolving a minor controversy in favor of
the numerical results of~\citet{Pappas:2013naa,Chakrabarti:2013tca}. 
Our results reproduce and extend the universal relations
of~\citep{I-Love-Q-Science,I-Love-Q-PRD,Pappas:2013naa}
to an arbitrary multipole number, although they deteriorate as $\ell$ grows.

Our results may point to a deep EoS universality between multipole moments of NSs,
which allow a description of their exterior gravitational field in terms of only three numbers.
Of course, not all NSs will present such universality, as we have here focused on the subset of 
old, cold, unmagnetized, and rigidly rotating stars.
However, all millisecond pulsar observations today and near-future GW detections 
are concerned with this subset.

Our work motivates the study of universality among multipole moments 
in the relativistic regime~\citep{NS-nohair}, which has a wide applicability in astrophysics, 
GWs, and experimental relativity. An observation of the NS mass, rotation period, and moment of inertia would suffice
to determine the first $7$ moments to within $10 \%$ accuracy,
irrespective of the NS EoS. In turn, the observations of several low multipole moments can in principle
be used to test GR irrespective of the NS EoS. Universal relations between multipole moments may be used to
break degeneracies in GW observations, probably with third-generation 
interferometers such as the Einstein Telescope~\citep{Punturo:2010zz}, increasing the accuracy of parameter extraction~\citep{I-Love-Q-PRD}.

Such universal relations may also be used in X-ray observations 
of NSs by breaking degeneracies in parameter extraction~\citep{baubock}.
For example, atomic features from NS surfaces like emission and absorption line profiles~\citep{2002Natur.420...51C} have already been used to
place constraints on the EoS~\citep{2006Natur.441.1115O}.
\citet{Baubock:2012bj} found that the quadrupole moment significantly affects the X-ray profile.
The spin of the fastest-spinning millisecond pulsar J1748+2446ad~\citep{2006Sci...311.1901H} can be as high as $\chi \sim 0.5$,
depending on its mass and the EoS. Given that the hexadecapole moment leads to corrections of roughly  
$\mathcal{O}(\chi^2)$ relative to the quadrupole moment, 
the former might not be negligible for rapidly rotating NSs. If so, the hexadecapole--quadrupole relation found here
should help to reduce the number of parameters and break degeneracies in X-ray observations. 

\emph{Acknowledgments}.~The authors acknowledge
K.~Chatziioannou, N.~Cornish, \'E.~Flanagan, J.~Lattimer, and T.~Tanaka for helpful discussions
and a careful reading of the manuscript. N.Y. and K.Y. acknowledge
support from NSF grant PHY-1114374 and the NSF CAREER Award
PHY-1250636, as well as support provided by the National Aeronautics
and Space Administration from grant NNX11AI49G, under sub-award
00001944. L.C.S. acknowledges that support for this work was provided by
the National Aeronautics and Space Administration through the Einstein
Postdoctoral Fellowship Award Number PF2-130101 issued by the Chandra
X-ray Observatory Center, which is operated by the Smithsonian
Astrophysical Observatory for and on behalf of the National
Aeronautics Space Administration under contract NAS8-03060, and
further acknowledges support from NSF grant PHY-1068541.

\bibliographystyle{apj}
\bibliography{master}
\end{document}